\documentclass[twocolumn,superscriptaddress,showpacs,amsmath,amssymb,aps,prb,floatfix,prstab,prstper,nolongbibliography]{revtex4-2}

\usepackage{color} 
\usepackage{graphicx}
\usepackage{braket} 
\usepackage{dcolumn}
\usepackage{bm}
\usepackage{subfigure}
\usepackage{amssymb}
\usepackage{multirow}
\usepackage{natbib}
\usepackage{amsmath}
\graphicspath{{plots/}}
\usepackage[english]{babel}

\newcommand{\add}[1]{{\color{black}#1}}

\usepackage[colorlinks,linkcolor=blue,anchorcolor=blue,citecolor=blue,urlcolor=blue,CJKbookmarks=True]{hyperref}

\begin{document}

\title{Towards Large-Scale and Spatio-temporally Resolved Diagnosis of \\Electronic Density of States \add{by Deep Learning}}

\author{Qiyu Zeng}
\author{Bo Chen}
\author{Xiaoxiang Yu}
\author{Shen Zhang}
\author{Dongdong Kang}
\affiliation{Department of Physics, National University of Defense Technology,Changsha, Hunan 410073, P. R. China}
\author{Han Wang}\email{wang\_han@iapcm.ac.cn} 
\affiliation{Laboratory of Computational Physics, Institute of Applied Physics and Computational Mathematics, Beijing 100088, P. R. China}
\author{Jiayu Dai} \email{jydai@nudt.edu.cn} 
\affiliation{Department of Physics, National University of Defense Technology,Changsha, Hunan 410073, P. R. China}

\date{\today}

%\begin{document}
\begin{abstract}
Modern laboratory techniques like ultrafast laser excitation and shock compression can bring matter into highly nonequilibrium states with complex structural transformation, metallization and dissociation dynamics. To understand and model the dramatic change of both electronic structures and ion dynamics during such dynamic processes, the traditional method faces difficulties. 
Here, we demonstrate the ability of deep neural network (DNN) to capture the atomic local-environment dependence of electronic density of states (DOS) for both multicomponent system under exoplanet thermodynamic condition and nonequilibrium system during super-heated melting process. Large scale and time-resolved diagnosis of DOS can be efficiently achieved within the accuracy of \textit{ab initio} method. Moreover, the atomic contribution to DOS given by DNN model accurately reveals the information of local neighborhood for selected atom, thus can serve as robust order parameters to identify different phases and intermediate local structures, strongly highlights the efficacy of this DNN model in studying dynamic processes.
\end{abstract}
% a shorter abstract
%\add{Here we demonstrate that our developed DeepDOS approach can not only serve as an efficient tool to study the temporal and spatial evolution of electronic density of state (DOS) during dynamic processes within the accuracy of \textit{ab initio} method, but also give robust identification for local structure that complicated by ionic thermal motion and phase transition dynamics. }

% keywords can be removed
\maketitle

%\textit{Introduction.}
\section{Introduction}
Modern laboratory techniques like ultrafast laser excitation and dynamic compression can bring matter into highly nonequilibrium states with complex structural transformation, metallization and dissociation dynamics\cite{zeng2020structural,chen2021atomistic,soubiran2018,boates2013}.
By increasing temperatures and pressures in a really short time scale ($ \sim 10^{-12}\ {\rm s}$), the behavior of both electrons and ions presents time-dependent characteristics.
Especially for the time-resolved diagnosis of electronic structures, it strongly suffers from the complication of phase transition associated with dramatic changes of thermodynamic state.
The demanding requirements for accurate description of electronic structure during dynamic processes raise challenges for all existing methods, and serve great importance for different communities like condensed matter physics, laboratory astrophysics, planetary physics, material science and industrial applications \cite{deringer2020,soubiran2018,jourdain2021}.

The classical molecular dynamics simulation (CMD), by integrating Newton's equations of motion for the many-body system, inherently describes the mass transport, energy transport and phase transition dynamics of ions with a large enough simulation size. However, the lack of explicit treatment of electrons prevents such a method from describing the evolution of electronic structure at the same time. While for \textit{ab initio} method based on quantum mechanics like density functional theory (DFT), it is too computational expensive to give the electronic response associated with long-time phase transition dynamics ($10^0\sim 10^3\ {\rm ps}$) and a large enough system ($>10^3$ atoms) that involve coexistence of different phases, gradients of temperatures and densities.

In past decades, there has been a burst of attempts to exploit machine-learning (ML) method to address challenges in the domain of molecular modeling. By directly approximating the solutions of the Kohn-Sham differential equations, ML models can efficiently yield interatomic potential energy surface \cite{behler2007, zhang2018deep} or electronic properties \cite{Chandrasekaran2019, Mahmoud2020, rio2020, Ellis2021} with high fidelity.
Most recently, ongoing efforts have demonstrated the usefulness of ML method on the electronic density \cite{Chandrasekaran2019}, electronic density of states (DOS) \cite{Mahmoud2020, rio2020} and the local density of states (LDOS)  \cite{Ellis2021}. Especially for the DOS, it serves as an important quantity for identification of metallization transition and \add{it is intimately related to properties like }electronic heat capacity, electron-phonon coupling and optical/X-ray absorption spectra \cite{lin2008, holst2014,ono2020ultrafast,jourdain2020understanding}.
Since the existing attempts have been restricted to \add{pure material} under ambient conditions, the ability of ML model to study \add{multicomponent system, the} temporal and spatial evolution of electronic properties during dynamic processes, has not been validated yet. 

\begin{figure*}[htbp]
\centering
\includegraphics[width=1\linewidth]{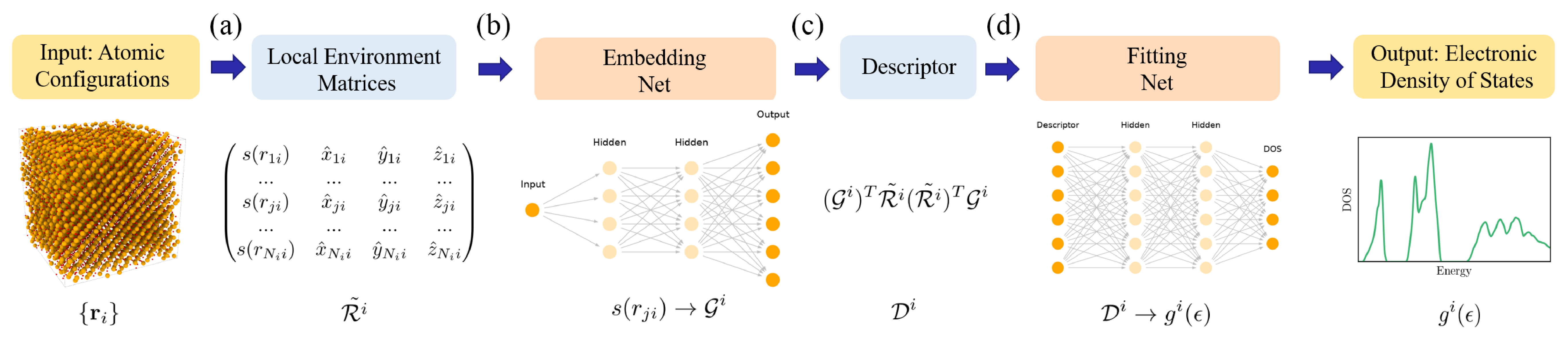}
\caption{Workflow of DeepDOS scheme. (a) Generating the local environmental matrices $\tilde {\mathcal R}^i$ from atomic configurations $\{\mathbf r_i\}$ to preserve the translation invariance. (b) Mapping smoothed relative distances $s(r_{ji})$ into $M$-dimension feature space to obtain local embedding matrices $\mathcal G^i$ via embedding net, where $s(...)$ the smooth function \cite{zhang2018end}. (c) Generating descriptor $\mathcal D^i$ by multiplication of local embedding matrices and local environmental matrices to preserve the permutation and rotation invariance. (d) Mapping the descriptor of atom $i$ into the atomic DOS $g^i(\epsilon)$ via fitting net.}
\label{fig:1}
\end{figure*}

Here, we utilize the deep neural network (DNN) to capture the atomic local environment dependence of DOS, named DeepDOS model.
The accuracy and robustness of DNN model is verified with magnesium oxide (MgO) system along the isotherm of 5000 kelvin, pressures ranging from 236 GPa to 598 GPa. Trained with small-system datasets, the DNN model is able to reproduce the DOS curves with high-fidelity for the unseen configurations generated by thermal fluctuation and density variation. 
Further, this size-extensive model is applied to efficiently study the temporal evolution of DOS curves during dynamics processes within the accuracy of \textit{ab initio} method. The isobaric melting process of silicon crystalline at pressure of 5 GPa is chosen. With a large enough system, the coexistence of solid and liquid phase can be incorporated and the corresponding DOS is given. Moreover, the atomic contribution to DOS accurately reveals the local structure for any selected atoms in this non-equilibrium system, and the DNN-predicted site-projected DOS can be used as fingerprint to characterize different local structures. This DeepDOS approach provides an efficient tool to study both the electronic structure and ionic local structure during dynamic processes created by shock impact, ultrafast laser heating etc.

\section{Method}
%\textit{Method. }
The machine-learning scheme for DOS prediction based on the two key approximation \cite{Mahmoud2020}:
(1) the DOS for a configuration can be decomposed into ``atomic contributions'' so as to preserve the extensive property. 
(2) each ``atomic contribution'' depends on the local environment of the atom, specifically, the relative coordinates of its neighbors within a pre-defined cut-off radius $r_{cut}$. The DNN-DOS model can be defined as
\begin{align}
g(\epsilon) =\sum_i \mathcal N_{\alpha_i}(\mathcal D_{\alpha_i}(r_i,\{r_j\}_{j\in n(i)}))
\end{align}
where $\mathcal N_{\alpha_i}$ denotes the neural network for specified chemical species $\alpha_i$ of atom $i$, and the descriptor $\mathcal D_{\alpha_i}$ describes the local environment of atom $i$ with its neighbor list $n(i) =\{j|r_{ji}<r_{cut}\}$.
Here we adopted the symmetry-preserving scheme introduced by DeepPot-SE model \cite{zhang2018end, wang2018deepmd, zhang2018deep} to generate the descriptor. 

As illustrated in the Fig.\ref{fig:1}, the sub-network consists of an embedding and a fitting neural network. 
The embedding network is specially designed to map the local environment $\tilde {\mathcal R^i}$ to an embedded feature space, which preserves the translational, rotational, and permutational symmetries of the system. The fitting network is a
fairly standard fully-connected feedforward neural network, which maps the descriptor to the atomic contribution to total DOS, denoted as atomic DOS (ADOS). 

 \add{To determine local electronic characteristics, the atomic contribution to DOS serves the key role. In traditional method, the total DOS is usually used to minimize the loss function, and decomposed into ADOS automatically by the machine learning model \cite{Mahmoud2020, rio2020}. Although the total DOS can be well reproduced, the predicted atomic contribution maybe inaccurate because only the sum of ADOS is constrained \cite{supple}. Here, to accurately desribe the different atomic contribution originated from different element type and atomic local environment, the site-projected DOS (PDOS) is used to minimize the loss function. Thus a loss function different with previous works is adopted to optimize the full parameters $\mathbf w$ of the DNN,} defined as the root mean square error (RMSE) of ADOS between model prediction $g^i_{pred,\mathbf w}$ and dataset $g^i(\epsilon)$.
\begin{align}
\mathcal L(\mathbf w)  =\sqrt{ \frac{1}{N} \sum_i^N \int_{\epsilon} (g^i_{pred,\mathbf w}(\epsilon)-g^i(\epsilon))^2 d\epsilon} \end{align}

\section{Dataset}
%\textit{Dataset.}
To generate the dataset, we perform the Deep Potential MD (DPMD) simulations to generate trajectories for two different systems with the LAMMPS package \cite{plimpton1995fast}, and label the configurations along the trajectories.

\begin{itemize}
\item Magnesium oxide (MgO, B1 phase) under exoplanet conditions at temperature of 5000 kelvin, pressures ranging from 236 GPa to 598 GPa.
\item Heat-until-melt simulations for silicon (cubic diamond phase) under isobaric conditions at pressure of 5 GPa. 
\end{itemize}

\begin{figure}[ht]
\centering
\includegraphics[width=1\linewidth]{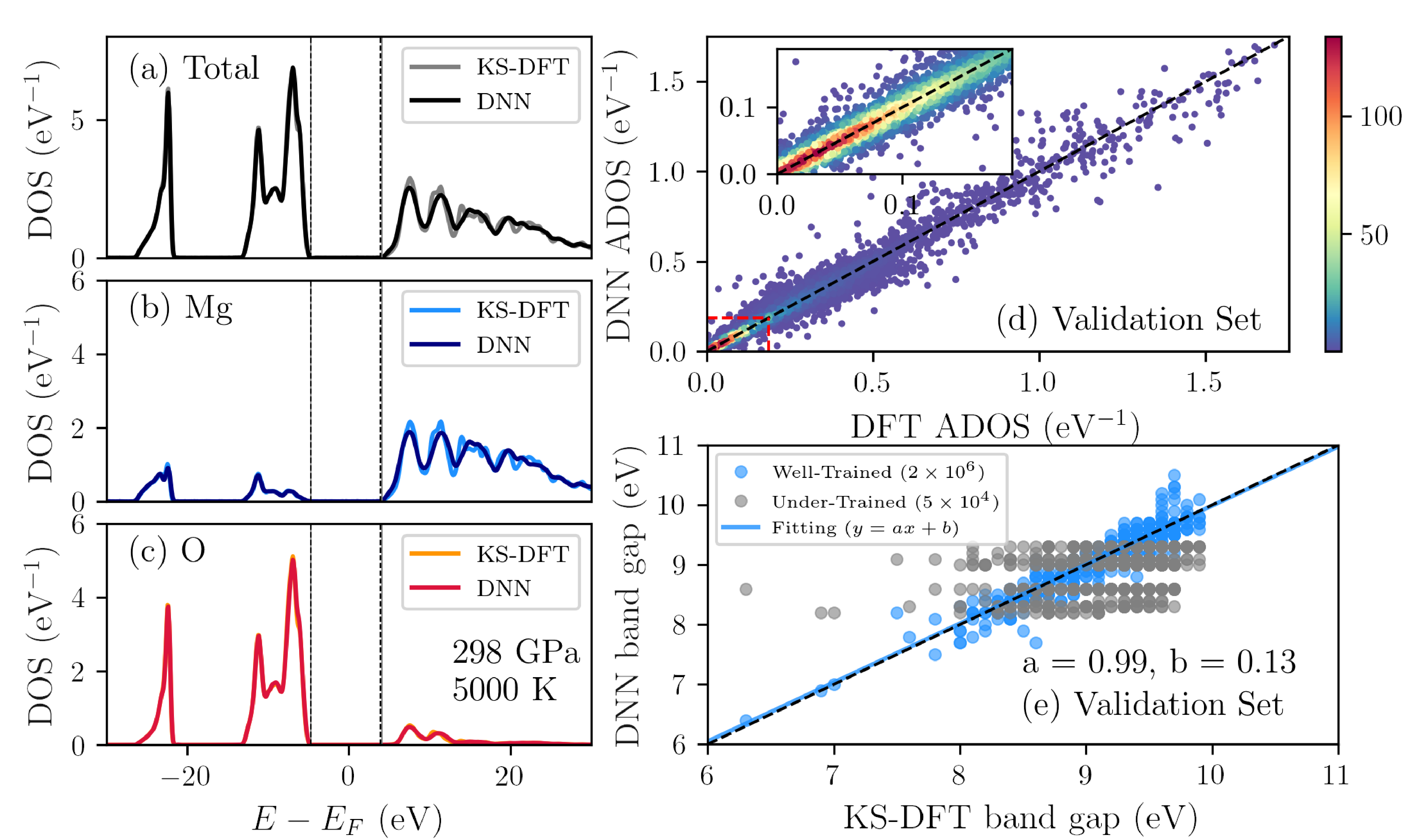}
\caption{(a)(b)(c) Comparison of DOS/ADOS between KS-DFT calculation and DNN prediction at $p=298\ {\rm GPa}$, $T=5000\ {\rm K}$. (d)(e) Parity plot of ADOS/width of band gap between DFT and DNN prediction in the validation set, where five thermodynamic conditions are included. In (d), the marker color indicates the frequency.  In (e), the blue and gray dots represent the predictions of band gap by well-trained DNN model (training steps of $2\times 10^6$) and under-trained DNN model ($5\times 10^4$) respectively. }
\label{fig:2}
\end{figure}

The DP models used in the DPMD simulations for MgO and Si are generated with DeePMD-kit packages \cite{wang2018deepmd}. To cover a wide range of thermodynamics conditions and minimize the computational consumption, a concurrent learning scheme, Deep Potential Generator (DP-GEN) \cite{zhang2019active}, has been adopted to sample the most compact and adequate data set that guarantees the uniform accuracy of DP in the explored configuration space. More details can be found in the supplemental material (SM) \cite{supple}.
In DPMD simulations, 40-ps-long molecular dynamics trajectories are generated with a timestep of 1.0 fs, and the snapshots are collected with an interval of 0.2 ps. To obtain the corresponding site-projected DOS along the trajectories, the QUANTUM ESPRESSO package \cite{giannozzi2017advanced} is used. The Perdew-Burke-Erzerhof (PBE) exchange correlation functional is used \cite{perdew1996generalized}, and the pseudopotential takes the projector augmented-wave (PAW) formalism \cite{blochl1994projector, holzwarth2001projector}. 
The sampling of Brillouin zone is chosen as 0.125 ${\rm \mathring A^{-1}}$. In the calculation of DOS, the Gaussian smearing is used and the width is chosen as 0.02 Ry for MgO and 0.01 Ry for silicon. To train a DeepDOS model, the
embedding network is composed of three layers (25, 50, and 100 nodes) while the fitting network has three hidden layers with 240 nodes in each layer. The radius cutoff $r_c$ is chosen to be $6.0\ {\rm \mathring A}$.

\section{Capturing thermal fluctuation and density variation}
%\textit{Capturing thermal fluctuation and density variation.} 
For a finite-temperature system, the ionic thermal motion plays an important role that drives thermal fluctuations both in the atomic local environment and electronic structure. 
Along the isotherm of MgO system at 5000 kelvin, ionic thermal motion can introduce significant fluctuations in the $3p$ band of Mg atoms, $2s$ and $2p$ band of oxygen atoms, indicating that both Mg-ADOS and O-ADOS are sensitive to atomic local environment even at exoplanet pressures (see \add{Fig.S2} and \add{Fig.S3} in SM). The standard deviation of ADOS is $\sigma_1 = 0.136\ {\rm eV^{-1}}, \sigma_2 = 0.262\ {\rm eV^{-1}}$ for Mg atom and O atom respectively. Such finite temperature effect can also lead to a large deviation in determining the width of band gap, varying from $\sim 6\ {\rm eV}$ to $\sim 9\ {\rm eV}$.

\begin{figure}[ht]
\centering
\includegraphics[width=1\linewidth]{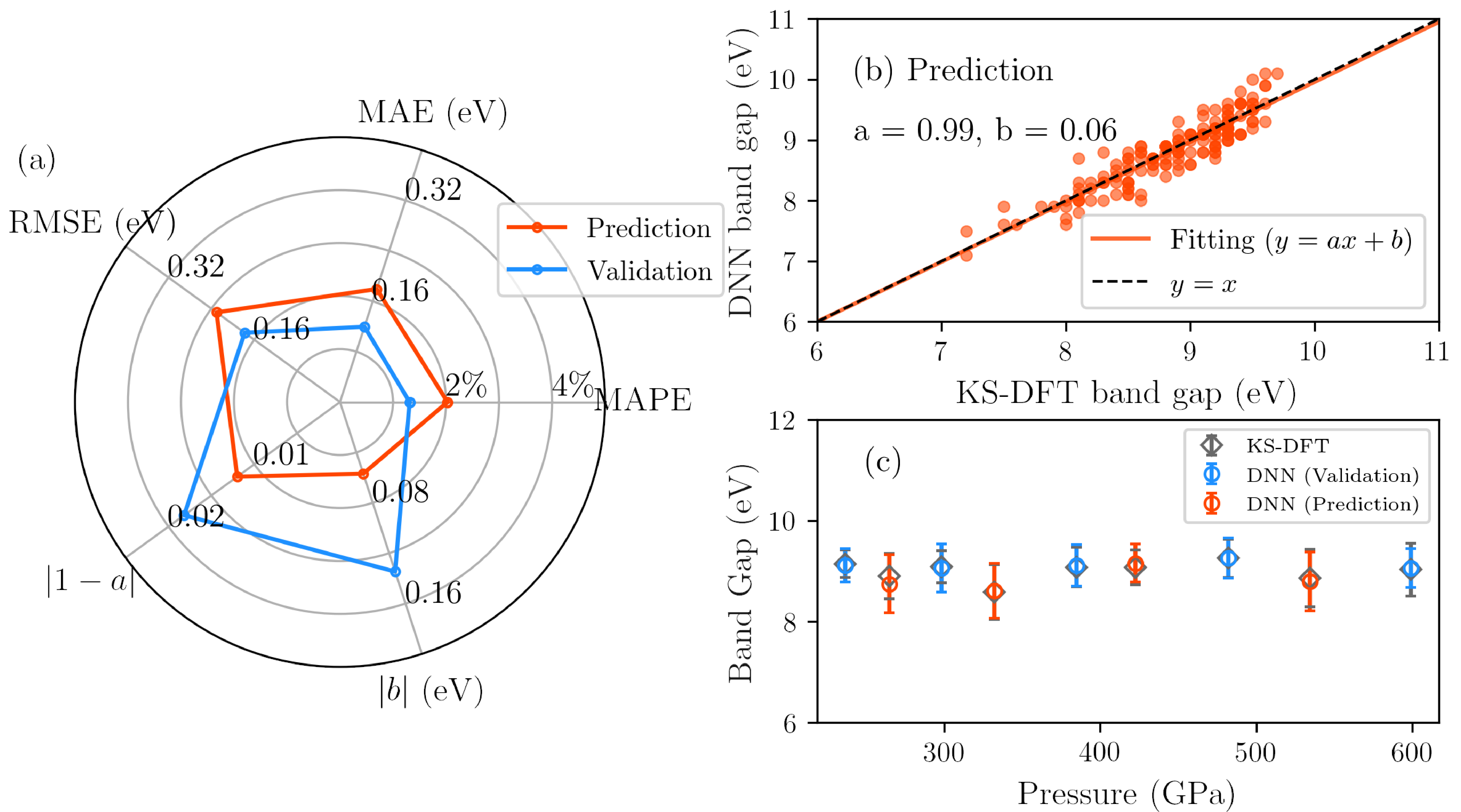}
\caption{(a) Radar plot of the test statistics chosen to measure the band gap prediction accuracy for validation set and test set. (b) DNN-Predicted band gaps versus KS-DFT band gaps on test set. Black dashed line is for visual reference, showing what the exact band gap prediction corresponds to. (c) Configuration-averaged band gaps at different pressures along the isotherm of 5000 kelvin, KS-DFT calculations (grey), DNN predictions on validation set (blue) and on test set (red) are presented. }
\label{fig:3}
\end{figure}

Here, we first test the performance of DNN model on DOS prediction for this multicomponent system with thermal fluctuated atomic configurations as input. 
To train our DNN model, we used a dataset covering five different exoplanet pressures from 237 GPa to 599 GPa, with each case containing 400 magnesium site-projected DOS and 400 oxygen site-projected DOS. This dataset is randomly divided into training set and validation set by ratio of 4:1.
As presented in the Fig.\ref{fig:2}(a)(b)(c), a well-trained DNN model can accurately reproduce the KS-DFT results of both the total DOS and site-projected DOS for an unseen atomic configuration. Fig.\ref{fig:2}(d) highlights the consistency between DNN model and \textit{ab initio} method in predicting the atomic contribution to the number of electronic states $g^i(\epsilon)$ at specific energy interval $\epsilon + d\epsilon$. 
The RMSE of Mg-ADOS and O-ADOS prediction are $0.115\sigma_1$ and $0.059\sigma_2$ respectively, significantly lower than the standard deviation of dataset. 

Moreover, the width of band gap is extracted from the predicted DOS curve, defined as the interval between valence band maximum (VBM) and conduction band minimum (CBM). 
As presented in the Fig.\ref{fig:2}(e), the alignment of the band gap predictions along the diagonal exhibits small standard deviations for thermal fluctuated atomic configurations.
The RMSE on the validation set is $0.17\ {\rm eV}$, the mean absolute error (MAE) is $0.11\ {\rm eV}$ and the mean absolute percentage error (MAPE) is $1.3\%$. The model performance on validation set indicates that the DeepDOS model has the ability to capture the changes of atomic contribution to electronic structure induced by thermal fluctuation of local environment, not only the shape of DOS curve, but also the detailed structure like the position of VBM and CBM.

\begin{figure}[ht]
\centering
\includegraphics[width=1\linewidth]{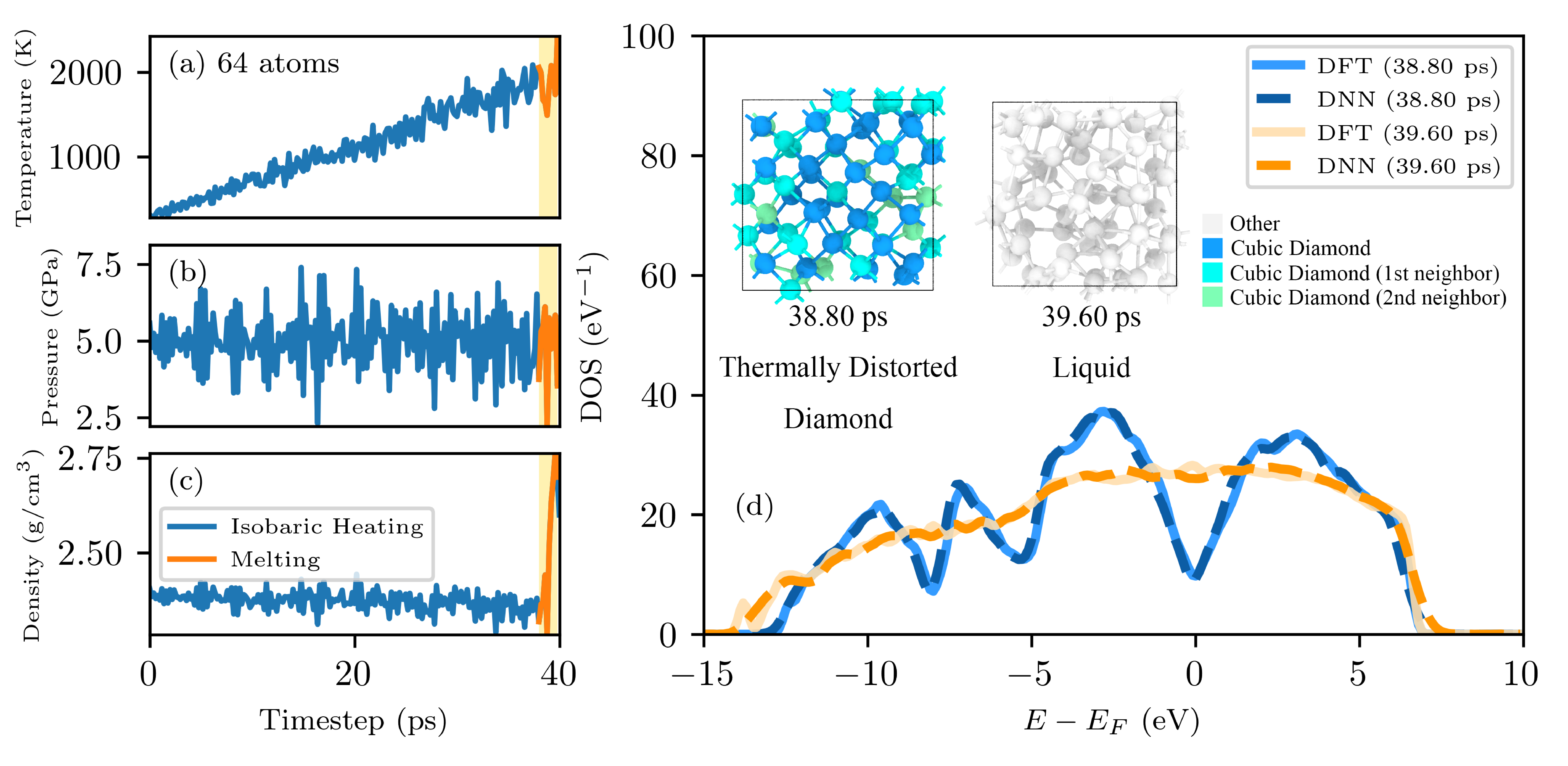}
\caption{(a)(b)(c) Temporal evolution of temperature, pressure and density during isobaric heating of cubic diamond silicon at 5 GPa (containing 64 atoms atoms), the orange line indicates the onset of melting. (d) Electronic density of states prediction for the unseen snapshots during isobaric melting process, the solid (dashed) line indicates result from KS-DFT calculation (DNN-prediction). The inset shows the snapshots of the atomic configurations colored by ext-CNA method \cite{maras2016}.}
\label{fig:4}
\end{figure}

For comparison, prediction given by an under-trained model (trained with $5\times 10^4$ steps) is presented in Fig.\ref{fig:2}(e). Interestingly, for a given thermodynamic condition, the under-trained model failed to describe the subtle difference of electronic states between thermal-fluctuated atomic configurations, leading to several discrete band gap prediction like a classification model for different thermodynamic conditions.

Further, the variation of density is considered. Since the inhomogeneous density distribution and temperature distribution can be created during dynamic processes, the ability for DNN to predict the unseen configurations for different densities not included in the training set, should be checked.

Four interpolated density points within the pressure range of training data are chosen to generate the dataset, denoted as the test set. As presented in the Fig.\ref{fig:3}(a), the performance of DeepDOS model in both validation set and test set show similar accuracy. In test set, the RMSE of band gap prediction is $0.23\ {\rm eV}$, the MAE is $0.18\ {\rm eV}$ and the MAPE is $2.0\%$. 
The DNN-predicted band gaps in Fig.\ref{fig:3}(b) are centered around the correct values, thus providing a low bias estimation. The predicted band gaps vs the calculated values from KS-DFT can be fitted to a linear function $y =ax+b$ with the slope $a = 0.99$ and bias $b= 0.06\ {\rm eV}$. 
The configuration-averaged band gap at different pressures are presented in Fig.\ref{fig:3}(c), the DNN exhibits a good generalization performance along the isotherm with pressure-sampling interval of 50 GPa, thus can provide a robust prediction in the study of dynamic processes. 

%(Results-2)
\section{Large-Scale and Time-Resolved DOS Prediction}
%\textit{Large-Scale and Time-Resolved DOS Prediction.} 

\begin{figure}[ht]
\centering
\includegraphics[width=1\linewidth]{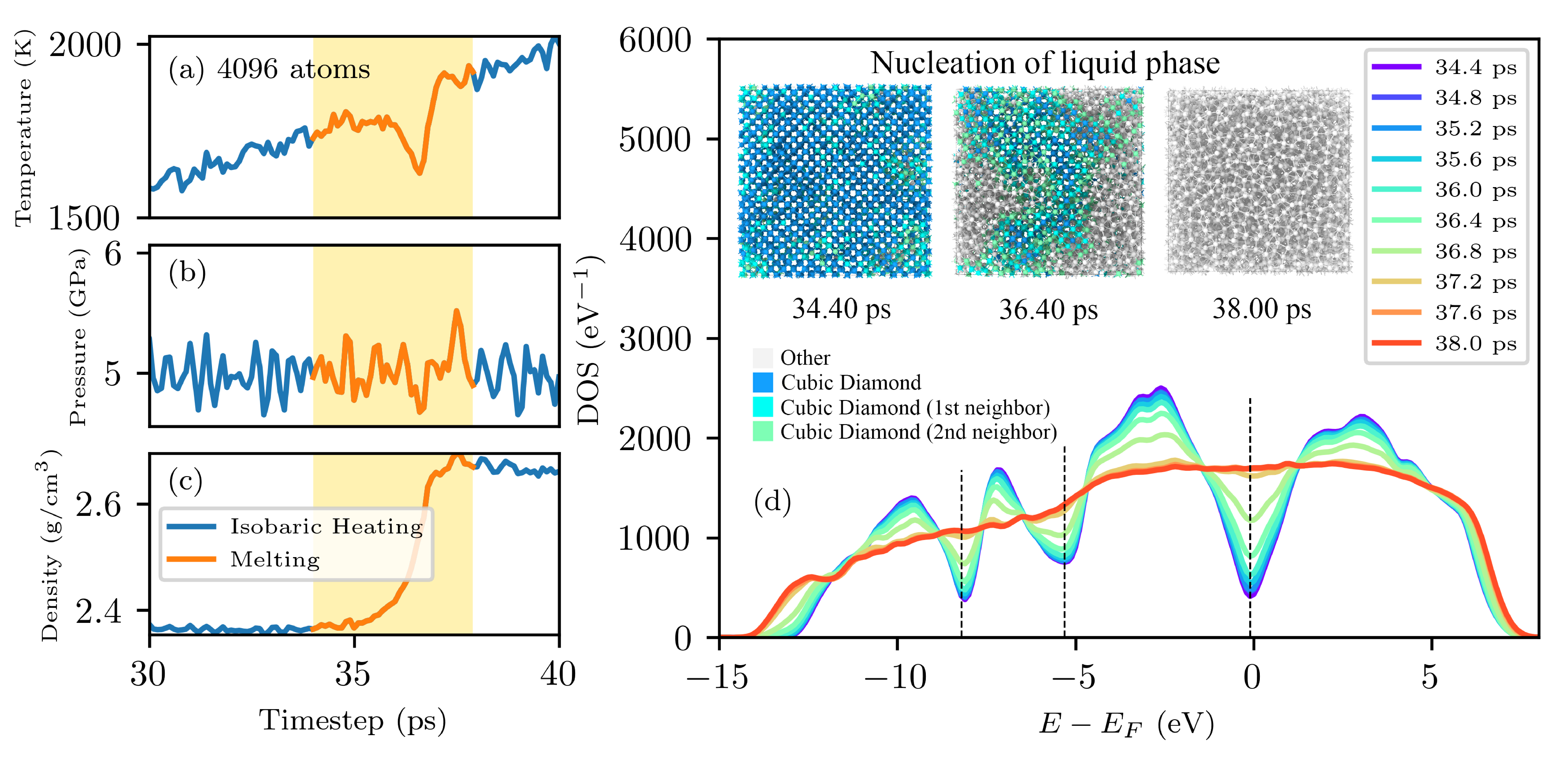}
\caption{(a)(b)(c) Temporal evolution of temperature, pressure and density during isobaric heating of cubic diamond silicon at 5 GPa (containing 4096 atoms atoms). (d) DOS prediction for the unseen snapshots of solid-liquid coexistence phase during the process of nucleation and growth of liquid phase.\add{The black dotted lines mark the three local minimum to give the characteristic parameter $Q$.}}
\label{fig:5}
\end{figure}

By combining non-equilibrium molecular dynamics with DeepDOS model, we investigate the temporal evolution of DOS during the superheated melting process of silicon, which contains solid-liquid coexistence, density gradients and thermal-fluctuated local structure.

The dataset containing 1600 silicon site-projected DOS is prepared. The atomic configurations are collected from the heat-until-melt simulation under isobaric condition, where the cold lattice, thermal-distorted crystalline and liquid phase are included. Different system size is considered, we chose 40 atomic configurations for 8-atom-system, 20 configurations for 16-atom-system and 10 configurations for 32-atom-system. Moreover, 80 snapshots for 8-atom-system equilibrated at T= 2000 kelvin, p = 5 GPa are used to improve the description of disordered structure.

%(64-atom prediction)
We validate our DeepDOS model by studying the melting process performed with 64 atom-system, where KS-DFT calculation is affordable.
We use DPMD to generate the trajectories for 64-atom system, and label the configurations along the trajectories by KS-DFT to provide the test set.
As shown in Fig.\ref{fig:4}(a)(b)(c), by heating silicon system from 300 kelvin to 2000 kelvin at 5 GPa, the cubic diamond structure maintains until 38.80 ps. 
As the temperature excesses the limit of thermal stability, the crystal collapses into disordered structure immediately accompanied with dramatic change of volume. Correspondingly, the abrupt changes of DOS curve induced by solid-liquid transition can be observed from Fig.\ref{fig:4}(d), where DNN predictions show good agreement with KS-DFT calculations for these unseen snapshots of 64-atom-system. Trained with small-system datasets, the DNN model is size extensive to reproduce the DOS curves with high-fidelity for large system.

%(4096-atom prediction)
Although the superheated melting is known to be proceeded by nucleation and growth of liquid regions inside the crystal. Due to limited simulation size, tens of or several hundred atoms are not enough to form the liquid cluster according to classic nucleation theory\cite{lu1998, rethfeld2002}. For removal of such size effect, simulations is performed with 4096-atom-system, and the DOS evolution show totally different change compared with 64-atom results.

\begin{figure}[htbp]
\centering
\includegraphics[width=1\linewidth]{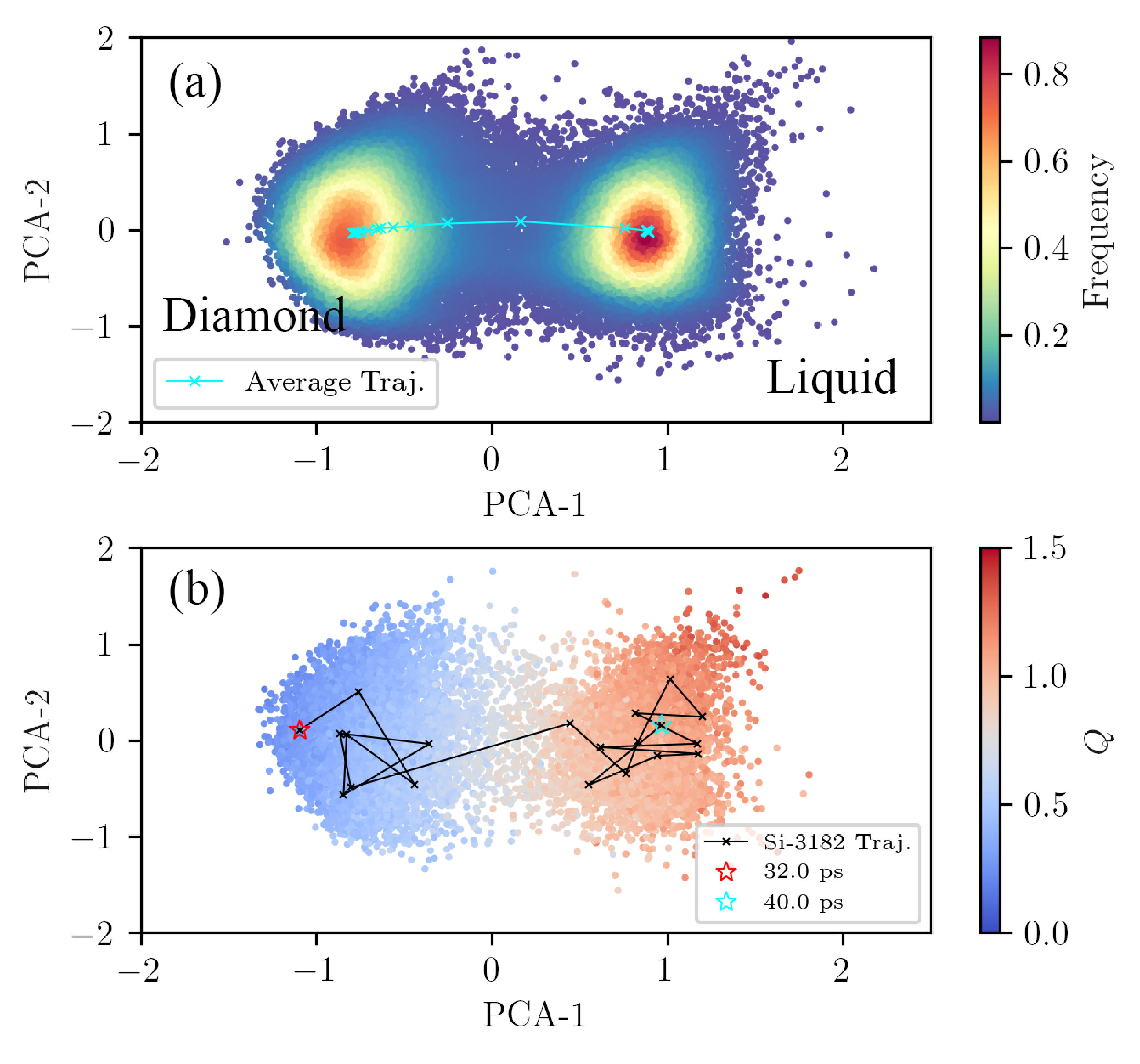}
\caption{(a)(b) PCA scatterplot of 81 920 ADOS from 20 snapshots of 4096-atom-system during melting process, where the marker color indicates the frequency and ADOS-derived order parameter $Q$ respectively. The cyan line in (a) indicates the trajectories of averaged ADOS in PCA space during melting process. In (b), a trajectory of Si-3182 atom in ADOS-PCA space is highlighted with black line.}
\label{fig:6}
\end{figure}

As presented in the insets of Fig.\ref{fig:5}(d),  when the temperature quickly rises above the normal melting point, melting occurs when the superheated crystal generates a sufficiently large number of destabilized particles \cite{jin2001melting}. Appearance and growth of small liquid regions inside the crystal proceeds the melting process.  
From 35.00 ps to 37.80 ps, the system is coexisted with solid and liquid phase, and well-defined crystal-liquid interfaces can be observed in a snapshot at 36.40 ps, identified by an extension of the common neighbor analysis method (ext-CNA) \cite{maras2016}. The behavior of electronic structure thus exhibits a hybrid characteristics of both phases as shown in Fig.\ref{fig:5}(d). During the coalescence of the liquid regions, we can observe a gradual change of DOS from solid-like semi-metal phase to liquid-like metal phase within $\sim 2.8$ ps, different from the abrupt change in 64-atom system results.  
We note that these results can hardly be achieved by traditional \textit{ab initio} method, and the DeepDOS model provides an opportunity to study the electronic structure for non-equilibrium processes, where both long-time ion dynamics and large system size are needed.

\section{Characterizing the atomic contribution to total DOS}
%\textit{Characterizing the atomic contribution to total DOS.} 

\begin{figure}[htbp]
\centering
\includegraphics[width=1\linewidth]{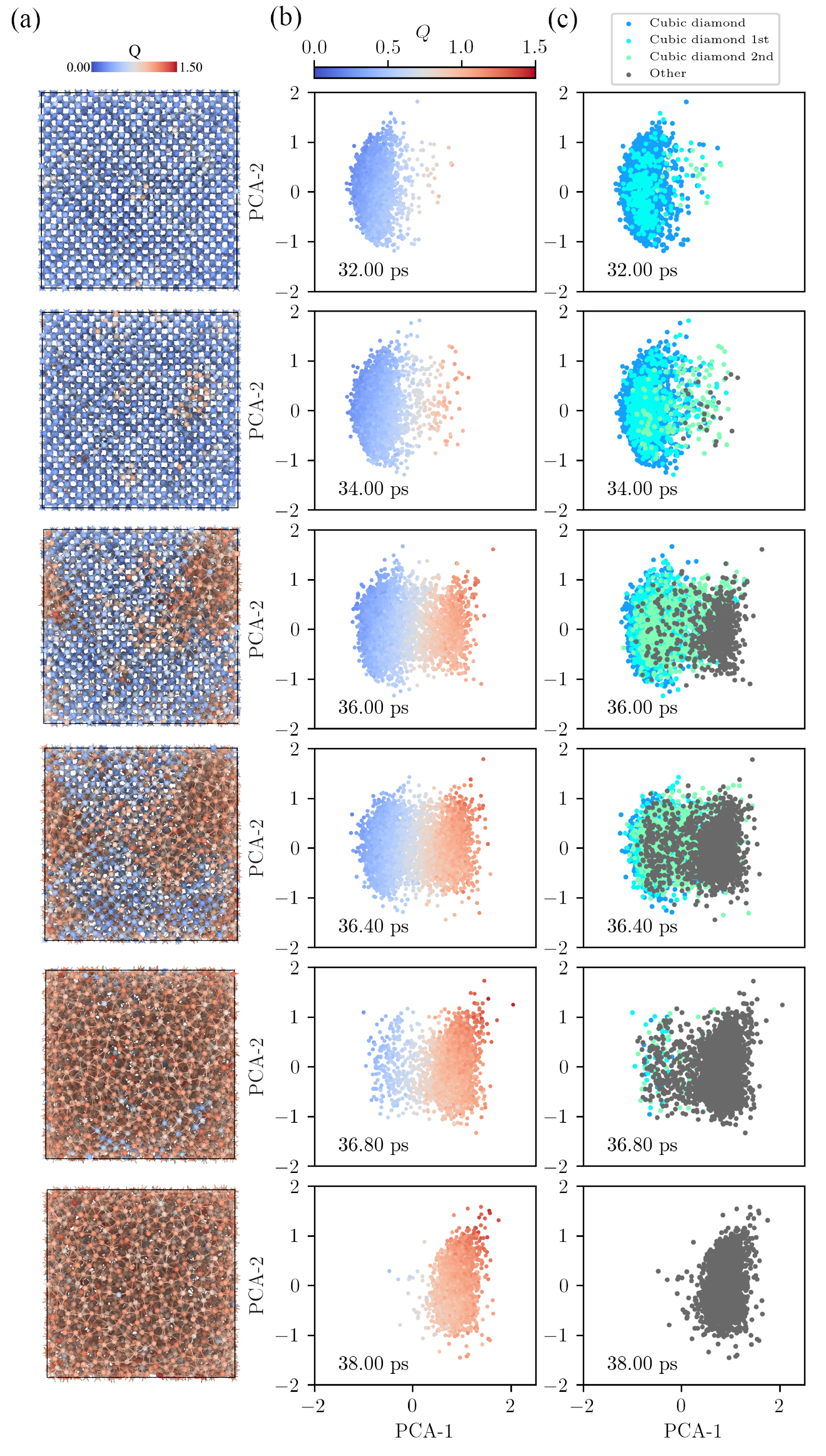}
\caption{(a) Snapshots of atomic configurations, colored by $Q$. (b)(c) Temporal evolution of ADOS in PCA subspace colored by ADOS-derived order parameter $Q$ and ext-CNA method respectively.  }
\label{fig:7}
\end{figure}
Based on the parameterized high-dimensional mapping between local environment and atomic contribution to total DOS, here we further investigate the temporal evolution of ADOS during melting process.
By applying principal component analysis (PCA), 81 920 ADOS extracted from 20 snapshots of 4096-atom-system during melting processes (from 34.0 ps to 40.0 ps) are projected onto two-dimensional principal component space.

As shown in Fig.\ref{fig:6}(a), due to different characteristics in the ADOS curves, the distribution of ADOS are separated into two clusters corresponding to the diamond phase and liquid phase, while the transitional region between two phases represents the intermediate local structure like strongly-distorted crystalline generated during melting processes. The electronic feature of these intermediate structures exhibits a combination of liquid and solid phase, as reported in previous section.

Choosing three electronic states from local minimum in DOS at Fermi level $E_F$, $\epsilon_1 \sim E_F-8\ {\rm eV}$ and $\epsilon_2 \sim E_F-5\ {\rm eV}$, a characteristic parameter $Q = g(E_F) + g(\epsilon_1)+g(\epsilon_2)$ for ADOS curves is given to describe the degree of disordered local environment for a selected atom. As presented in Fig. \ref{fig:6}(b), the characteristic parameter gradually increases from $Q_{min}\sim 0.19$ to $Q_{max}\sim 1.46$ as the representative point in PCA subspace flows from diamond-region into liquid-region, indicating that such ADOS-derived quantity can serve as a robust order parameter to identify the liquid phase, diamond phase and intermediate structure.

The temporal evolution of ADOS during melting process is presented in the Fig.\ref{fig:7}(b). Before the onset of melting at 32.0 ps, the representative points in PCA subspace concentrates to form the diamond-phase island. As melting occurs,
with the coalescence of liquid region and the movement of liquid-solid interface, these representative points gradually flow away from the diamond region, and finally settle down in the liquid-phase region, show good consistency with the snapshots of atomic configurations in the Fig.\ref{fig:7}(a). We note that a significant amount of representative points locate in the transitional zone from 34.0 ps to 36.8 ps, with characteristic parameter $Q$ ranging from $\sim 0.6$ to $\sim 0.8$, indicating that the DNN-predicted ADOS and ADOS-derived order parameter $Q$ accurately resolves the complex local environment for this nonequilibrium system that contains diamond crystalline, liquid and thermal-distorted intermediate structure.

For comparison, the traditional structure-identification method like ext-CNA method is used to classify the local structure. Based on the geometric feature of atomic local neighborhoods for selected atoms, the ext-CNA method is sensitive to the structural noise driven by ionic thermal motion and phase transition dynamics, thus failed to classify all these local structures correctly. As presented in the Fig.\ref{fig:7}(c), the distributions of pre-defined reference structures in the PCA subspace all overlap one another.
These results indicate that the local electronic characteristics can achieve more accurate and robust performance in identifying different phases compared with geometrical characteristics like coordination number and bond-orientational order parameter.

%\textit{Summary.} 
\section{Summary}
For \add{transient states widely exist during} dynamic processes, both the electronic structure and the atomic local neighborhoods can be strongly complicated by \add{extreme thermodynamic conditions, }inhomogeneity of  density distribution, coexistence and interface of different phases. Here we demonstrate that our developed DeepDOS approach can not only serve as an efficient tool to study the temporal-spatial evolution of DOS during dynamic processes within the accuracy of \textit{ab initio} method, but also provide a possibility to distinguish different phases and identify hidden patterns through local electronic features.

Specifically, the accuracy and robustness of the DeepDOS model are validated by reproducing DOS curves and band gaps for unseen atomic configurations generated by thermal fluctuation and density variations. Trained with small-system dataset, this size extensive model is able to give DOS curves for a large system. Therefore, time-resolved diagnosis of DOS under isobaric heating condition can be efficiently achieved by this machine-learning method, and we present the temporal evolution of DOS curves during melting processes that proceed by nucleation and growth of liquid region out of the solid matrix.
Moreover, the electronic feature for local structures can also be accurately captured. As compared with geometric characteristics like ext-CNA method, the ADOS can give relatively robust identification for local structures that significantly deviates from ideal crystal symmetry due to thermal fluctuation and phase transition dynamics.

In conclusion, this DeepDOS model can strongly enhance the further study of both electronic structure and ion dynamics during non-equilibrium processes created by ultrafast laser heating, shock loading or a quasi-isentropic ramped pressure drive.

%exhibits a better performance than geometrical feature to distinguish different phases and identify hidden patterns, providing a possibility for structure identification of complex system.

\section{Acknowledgments}
This work was supported by the National Key R\&D Program of China under Grant No. 2017YFA0403200, the National Natural Science Foundation of China under Grant No. 11774429, 11874424, 11904401, 12047561, 12104507, the NSAF under Grant No. U1830206, the Science and Technology Innovation Program of Hunan Province under Grant No. 2020RC2038, 2021RC4026.

\bibliography{refs} %%% Remove comment to use the external .bib file (using bibtex).


% George Kour and Raid Saabne.
% \newblock Real-time segmentation of on-line handwritten arabic script.
% \newblock In {\em Frontiers in Handwriting Recognition (ICFHR), 2014 14th
% International Conference on}, pages 417--422. IEEE, 2014.

% George Kour and Raid Saabne.
% \newblock Fast classification of handwritten on-line arabic characters.
% \newblock In {\em Soft Computing and Pattern Recognition (SoCPaR), 2014 6th
% International Conference of}, pages 312--318. IEEE, 2014.

% \bibitem{hadash2018estimate}
% Guy Hadash, Einat Kermany, Boaz Carmeli, Ofer Lavi, George Kour, and Alon
% Jacovi.
% \newblock Estimate and replace: A novel approach to integrating deep neural
% networks with existing applications.
% \newblock {\em arXiv preprint arXiv:1804.09028}, 2018.

% \end{thebibliography}

\end{document}